# Crackling noise

James P. Sethna*, Karin A. Dahmen†, and Christopher R. Myers¶

*Laboratory of Atomic and Solid State Physics, Clark Hall, Cornell University, Ithaca, NY 14853-2501, sethna@lassp.cornell.edu. † Department of Physics, 1110 West Green Street, University of Illinois at Urbana-Champaign, IL 61801-3080, dahmen@physics.uiuc.edu. ¶ Cornell Theory Center, Frank H.T. Rhodes Hall, Cornell University, Ithaca, NY 14853-3801, myers@tc.cornell.edu.

**Crackling noise arises when a system responds to changing external conditions through discrete, impulsive events spanning a broad range of sizes. A wide variety of physical systems exhibiting crackling noise have been studied, from earthquakes on faults to paper crumpling. Because these systems exhibit regular behavior over many decades of sizes, their behavior is likely independent of microscopic and macroscopic details, and progress can be made by the use of very simple models. The fact that simple models and real systems can share the same behavior on a wide range of scales is called *universality*. We illustrate these ideas using results for our model of crackling noise in magnets, explaining the use of the *renormalization group* and *scaling collapses*. This field is still developing: we describe a number of continuing challenges.**

## Crackling noise: a new realm for science

In the past decade or so, science has broadened its purview to include a new range of phenomena. Based on advances in the 1970's on second-order phase transitions[0.5-4] and stochastic theories of turbulence[5] and in the 1980's on disordered systems,[6-8] we now claim that we should be able to explain how and why things crackle.

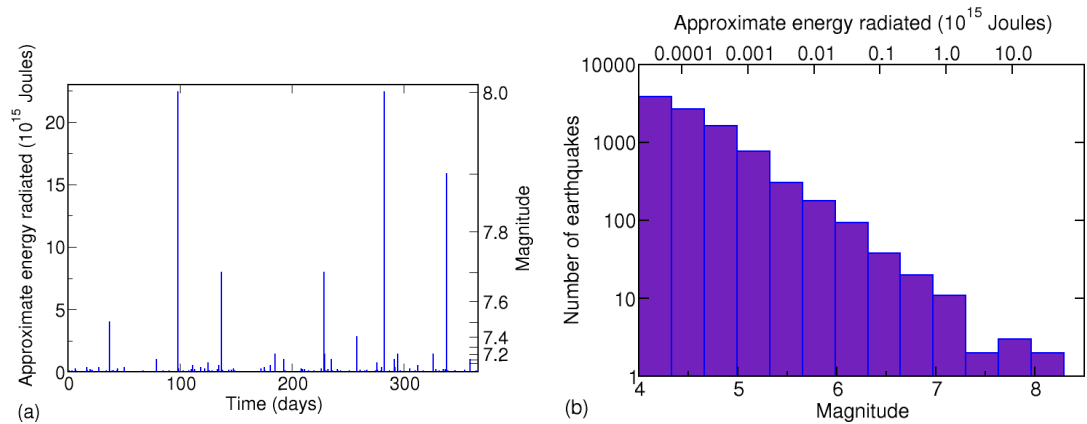

Figure 1: The Earth Crackles. (a) Time history of radiated energy from earthquakes throughout all of 1995.[10-12] The earth responds to the slow strains imposed by continental drift through a series of earthquakes: impulsive events well separated in space and time. This time series, when sped up, sounds remarkably like the crackling noise of paper, magnets, and Rice Krispies[©] (hear it in Ref. 12). (b) Histogram of number of earthquakes in 1995 as function of their magnitude (or, alternatively, their energy release). Earthquakes come in a wide range of sizes, from unnoticeable trembles to catastrophic events. The smaller earthquakes are much more common: the number of events of a given size forms a power law[9] called the Gutenberg-Richter law. (Earthquake magnitude scales with the logarithm of the strength of the earthquake, *e.g.* its radiated energy. On a log-log plot of number vs. radiated energy, a power law is a straight line, as we observe in the plotted histogram.) One would hope that such a simple law should have an elegant explanation.

Many systems crackle; when pushed slowly, they respond with discrete events of a variety of sizes. The earth responds[9] with violent and intermittent earthquakes as two tectonic plates rub past one another (see figure 1). A piece of paper[14] (or a candy wrapper at the movies[15,16]) emits intermittent, sharp noises as it is slowly crumpled or rumpled. (Try it: preferably not with this page.) A magnetic material in a changing external field magnetizes in a series of jumps.[13,17] These individual events span many orders of magnitude in size – indeed, the distribution of sizes forms a power law with no characteristic size scale. In the past few years, scientists have been making rapid progress in developing models and theories for understanding this sort of scale-invariant behavior in driven, nonlinear, dynamical systems.

Interest in these sorts of phenomena goes back several decades. The work of Gutenberg and Richter[9] in the 1940's and 1950's established the well-known frequency-magnitude relationship for earthquakes that bears their names (figure 1). A variety of many-degree-of-freedom dynamical models,[18-30] with and without disorder, have been introduced in the years since to investigate the nature of slip complexity in earthquakes. More recent impetus for work in this field came from the study of the depinning transition in sliding charge-density wave (CDW) conductors in the 1980's and early 1990's.[31-37] Interpretation of the CDW depinning transition as a dynamic critical phenomenon sprung from Fisher's early work,[31,32] and several theoretical and numerical studies followed. This activity culminated in the RG solution by Narayan and Fisher[34] and the numerical studies by Middleton[35] and Myers,[36] which combined to provide a clear picture of depinning in CDWs and open the doors to the study of other disordered, nonequilibrium systems.

Bak, Tang, and Wiesenfeld inspired much of the succeeding work on crackling noise.[38,39] They introduced the connection between dynamical critical phenomena and crackling noise, and they emphasized how systems may naturally end up at the critical point through a process of *self-organized criticality*. (Their original model was that of avalanches in growing sandpiles. Sand has long been used as an example of crackling noise.[40,41] However, it turns out that real sandpiles don't crackle at the longest scales.[42,43])

Researchers have studied many systems that crackle. Simple models have been developed to study bubbles rearranging in foams as they are sheared,[44] biological extinctions[45] (where the models are controversial:[46,47] of course we personally believe that the asteroid did in the dinosaurs), fluids invading porous materials and other problems involving invading fronts[48-53] (where the model we describe was invented[48,49]), the dynamics of superconductors[54-54.4] and superfluids,[55,56] sound emitted during martensitic phase transitions,[57] fluctuations in the stock market,[58,59] solar flares,[60] cascading failures in power grids,[61,62] failures in systems designed for optimal performance,[63-65] group decision making,[65.1] and fracture in disordered materials.[65.2-65.7] These models are driven systems with many degrees of freedom, which respond to the driving in a series of discrete avalanches spanning a broad range of scales – what we are calling crackling noise.

There has been healthy skepticism by some established professionals in these fields to the sometimes grandiose claims by newcomers proselytizing for an overarching paradigm. But often confusion arises because of the unusual kind of predictions the new methods provide. If our models apply at all to a physical system, they should be able to predict all behavior on long length and time scales, independent of many microscopic details of the real world. This predictive capacity comes, however, at a price: our models typically don't make clear predictions of how the real-world microscopic parameters affect the long-length-scale behavior.

In this paper, we will provide an overview of the *renormalization-group*[0.5-4] many researchers use to understand crackling noise. Briefly, the renormalization group discusses how the effective evolution laws of our system change as we measure on longer and longer length scales. (It works by generating a coarse-graining mapping in system space, the abstract space of all possible evolution laws.) The broad range of event sizes will be attributed to a self-similarity, where the evolution laws look the same under different length scales. Using this self-similarity, we are led to a method for scaling experimental data. In the simplest case this yields power laws and fractal structures, but more generally it leads to universal scaling functions – where we argue the real predictive power lies. We will only touch upon the dauntingly complex analytical methods used in this field, but we believe we can explain faithfully and fully both what our tools are useful for, and how to apply them in practice. The renormalization group is perhaps the most impressive use of abstraction in science.

**Why should crackling noise be comprehensible?**

Not all systems crackle. Some respond to external forces with lots of similar-sized, small events (popcorn popping as it is heated). Others give way in one single event (chalk snapping as it is stressed). Roughly speaking, crackling noise is in between these limits: when the connections between parts of the system are stronger than in popcorn but weaker than in the grains making up chalk, the yielding events can span many decades of sizes. Crackling forms the transition between snapping and popping.

Figure 1b presents a pretty simple relationship. We expect that there ought to be a simple, underlying reason that earthquakes occur on all different sizes. The very small earthquake properties probably depend a lot on the kind of dirt (fault gouge) in the crack. The very largest earthquakes will depend on the geography of the continental plates. But the smooth power-law behavior suggests that something simpler is happening in between, independent of either the microscopic or the macroscopic details.

There is a nice analogy with the behavior of a fluid. A fluid is very complicated on the microscopic scale, where molecules are bumping into one another: the trajectories of the molecules are chaotic, and depend both on exactly what direction they are moving and what they are made of. However, a simple law describes most fluids on long length and time scales. This law, the Navier-Stokes equation, depends on the constituent molecules only through a few parameters (the density and viscosity). Physics works because simple laws emerge on large scales. In fluids, these microscopic fluctuations and complexities go away on large scales: for crackling noise, they become scale invariant and self-similar.

How do we derive the laws for crackling noise? There are two approaches. First, one can analytically calculate the behavior on long length and time scales by formally coarse-graining over the microscopic fluctuations. This leads us to renormalization-group methods,[0.5-4] which we discuss in the next section. The analytic approach can get pretty hairy, but it can give useful results and (more importantly) is the only explanation for *why* events on all scales should occur. Second, one can make use of *universality*. If the microscopic details don't matter for the long length scale behavior, why not make up a simple model with the same behavior (in the same *universality class*) and solve it?

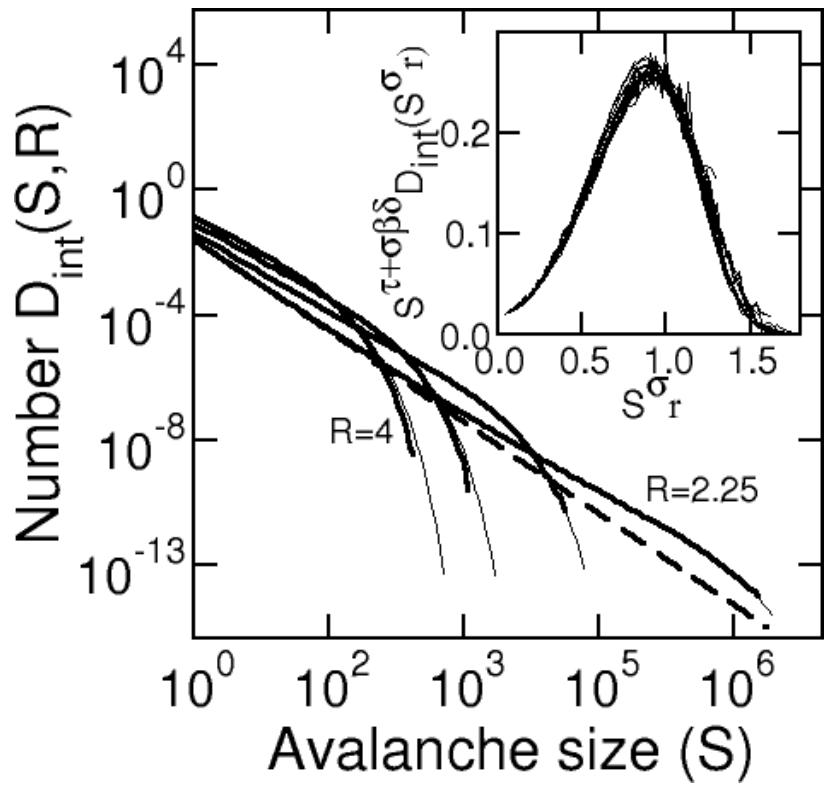

Figure 2: Magnets Crackle.[66-69] Magnets respond to a slowly varying external field by changing their magnetization in a series of bursts, or avalanches. These bursts, called Barkhausen noise, are very similar (albeit on very different time and size scales) to those shown in figure 1 for the earthquakes. The avalanches in our model have a power law distribution only at a special value of the disorder, $R_c$=2.16. Shown is the histogram giving the number of avalanches $D_{int}(S,R)$ of a given size $S$ at various disorders $R$ ranging from 4 to 2.25; the thin lines are theoretical predictions from our model. ($D_{int}$ gives all the avalanches during our simulation, integrated over the external field $-\infty < H(t) < +\infty$). The straight dashed line shows the power-law distribution at the critical point. Notice that fitting power laws to the data would work only very near to $R_c$: even with six decades of avalanche sizes, the slope hasn't converged to the asymptotic value. On the other hand, the scaling function predictions (theoretical curves) work well quite far from the critical point. The inset shows a scaling collapse of the avalanche size distribution (scaled probability versus scaled size), which is used to provide the theoretical curves as described in the text.

The model we'll focus on in this paper is a caricature of a magnetic material.[17,48,49,66-70] A piece of iron will "crackle" as it enters a strong magnetic field, giving what is called Barkhausen noise. We model the iron as a cubic grid of magnetic domains $S_i$, whose north pole is either pointing upwards ($S_i = +1$) or downward ($S_i = -1$). The external field pushes on our domain with a force $H(t)$, which will increase with time. Iron can be magnetized because neighboring domains prefer to point in the same direction: if the six neighbors of our cubic domain are $S_j$, then in our model we let their force on our domain be $\sum_j J\, S_j$. Finally, we model dirt, randomness in the domain shapes, and other kinds of disorder by introducing a random field $h_i$, different for each domain and chosen at random from a normal distribution with standard deviation $R$, which we call the disorder. The net force on our domain is thus

Force on domain $i = H(t) + \sum_j J\, S_j + h_i.$ [1]

The domains in our model all start pointing down (-1), and flip up as soon as the net force on them becomes positive. This can occur either because $H(t)$ increases sufficiently (spawning a new avalanche), or because one of their neighbors flipped up, kicking them over (propagating an existing avalanche). (Thermal fluctuations are ignored: a good approximation in many experiments because the domains are large.) If the disorder $R$ is large, so the $h_i$ are typically big compared to $J$, then most domains flip independently: all the avalanches are small, and one gets popping noise. If the disorder is small compared to $J$, then typically most of the domains will be triggered by one of their neighbors: one large avalanche will snap up most of our system. In between, we get crackling noise. When the disorder $R$ is just large enough so that each domain flip on average triggers one of its neighbors (at the critical disorder $R_c$), then we find avalanches on all scales (figures 2 and 3).

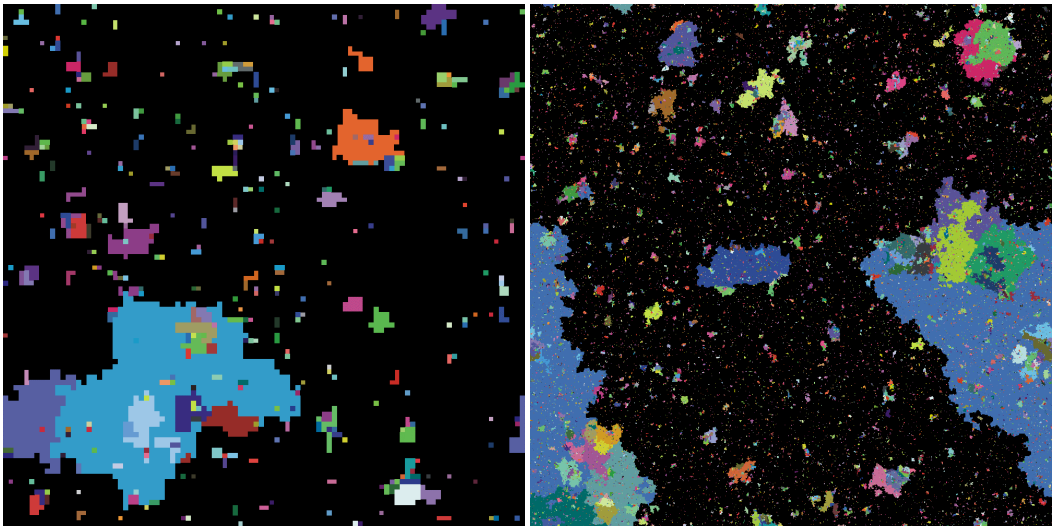

Figure 3: Self-similarity. These are cross-sections of the avalanches during the magnetization of our model.[17,66-69] Here each avalanche is drawn in a separate color. (a) shows a $100^3$ simulation and (b) shows a $1000^3$ simulation (a billion domains[21b]); both are run at the critical point $R_c$=2.16 $J$ where avalanches just barely continue. The black background represents a large avalanche that spans the system: the critical point occurs when avalanches would first span an infinite system.

What do these avalanches represent? In nonlinear systems with many degrees of freedom, there are often large numbers of metastable states. Local regions in the system can have multiple stable configurations, and many combinations of these local configurations are possible. (A state is metastable when it cannot lower its energy by small rearrangements. It's distinguished from the globally stable state, which is the absolute lowest energy possible for the system.) Avalanches are precisely the rearrangements as our system shifts from one metastable state to another. Our specific interest is in systems with a broad distribution of avalanche sizes, where shifting between metastable states can rearrange anything between a few domains and millions of domains.

There are lots of choices we made in our model that don't matter. Because of universality, we can argue[72-73] that the behavior would be the same if we chose a different grid of domains, or if we changed the distribution of random fields, or if we introduced more realistic random anisotropies and random coupling constants. Were this not the case, we could hardly expect our simple model to explain real experiments.

**The Renormalization Group and Scaling**

To study crackling noise, we use *renormalization-group*[0.5-4,71,72] tools developed in the study of second-order phase transitions. The word renormalization has roots in the study of quantum electrodynamics, where the effective charge changes in size (norm) as a function of length scale. The word group refers to the family of coarse-graining operations basic to the method: the group product is composition (coarsening repeatedly). The name is unfortunate, however, as the basic coarse-graining operation does not have an inverse, and thus the renormalization group does not have the mathematical structure of a group.

The renormalization group studies the way the space of all physical systems maps into itself under coarse-graining (see figure 4). The coarse-graining operation shrinks the system, and removes degrees of freedom on short length scales. Under coarse-graining, we often find a fixed point **S\***: many different models flow into the fixed point and hence share long-wavelength properties. To get a schematic view of coarse-graining, look at figure 3: the $1000^3$ cross section looks (statistically) like the $100^3$ section if you blur your eyes by a factor of 10. Much of the mathematical complexity of this field involves finding analytical tools for computing the flow diagram in figure 4. Using methods developed to study thermodynamical phase transitions[2] and the depinning of charge-density waves,[34] we can calculate for our model the flows for systems in dimensions close to six (the so-called $\varepsilon$ expansion,[71-73] where $\varepsilon$ =6-$d$, $d$ being the dimension of the system). Interpolating between dimensions may seem a surprising thing to do. In our system it gives rather good predictions even in three dimensions (i.e., $\varepsilon$=3), but it's hard work, and we won't discuss it here. Nor will we discuss real-space renormalization-group methods[0.5] or series expansion methods. We focus on the relatively simple task of using the renormalization group to justify and explain the universality, self-similarity, and scaling observed in nature.

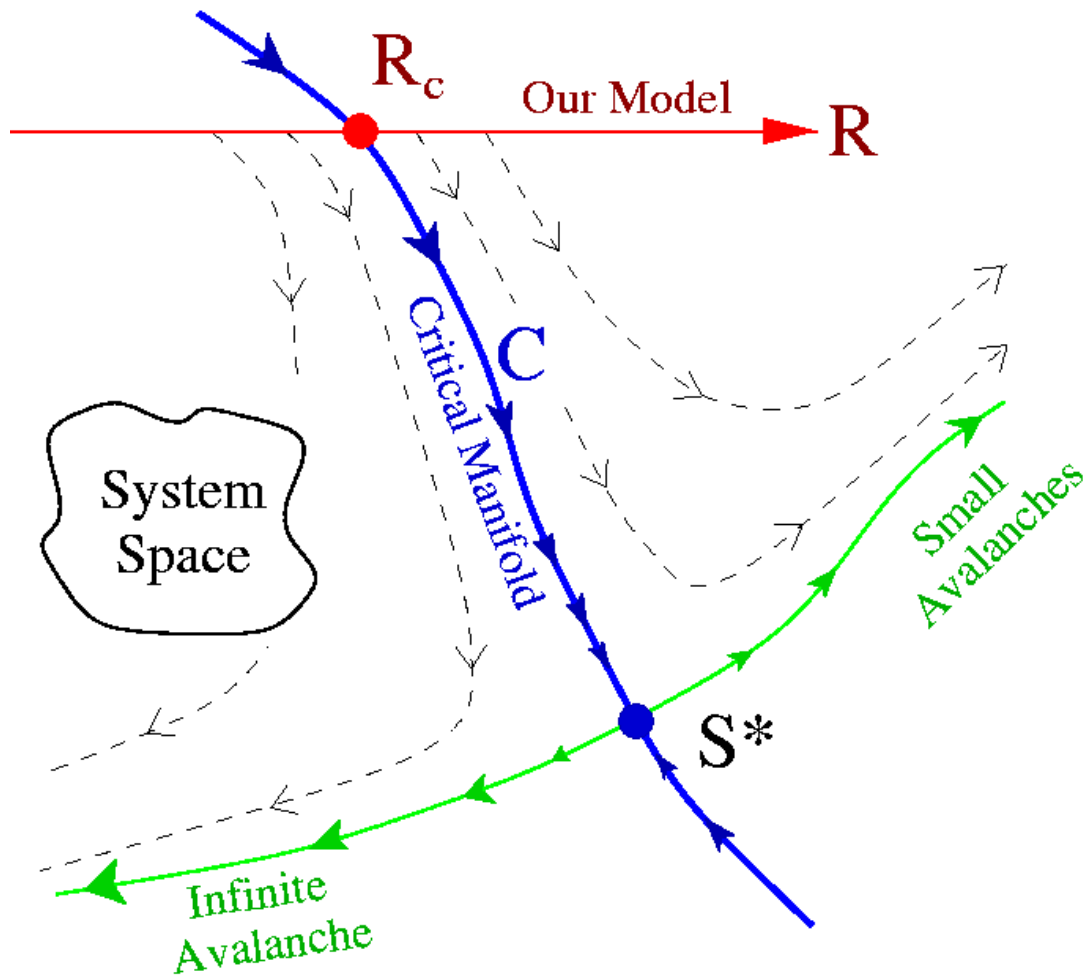

Figure 4: Renormalization-group flows. The renormalization-group is a theory of how coarse-graining to longer length scales introduces a mapping from the space of physical systems to itself. Consider the space of all possible models of magnetic hysteresis (what an abstraction!). Each model can be coarse-grained, removing some fraction of the microscopic degrees of freedom and introducing more complicated rules so that the remaining ones still flip at the same external fields. This defines a mapping from our space into itself. A fixed point S* in this space will be self-similar: since it maps to itself upon coarse-graining, it must have the same behavior on different length scales. Points that flow into S* under coarse-graining share this self-similar behavior on sufficiently long length scales: they all share the same *universality class*.

Consider the "system space" for disordered magnets. There is a separate dimension in system space for each possible parameter in a theoretical model (disorder, coupling, next-neighbor coupling, dipolar fields ...) or in an experiment (temperature, annealing time, chemical composition ...). Coarse-graining, however one implements it, gives a mapping from system space into itself: shrinking the system and ignoring the shortest length scales yields a new physical system with identical long-distance physics, but with different (renormalized) values of the parameters. We've abstracted the problem of understanding crackling noise in magnets into understanding a dynamical system acting on the space of all dynamical systems.

Figure 4 represents a two-dimensional cross section of this infinite-dimensional system space. We've chosen the cross section to include our model (equation [1]): as we vary the disorder $R$, our model sweeps out a straight line (red) in system space. The cross section also includes a fixed point **S\***, which maps into itself under coarse-graining. The system **S\*** looks the same on all length and time scales, because it coarse-grains into itself. We can picture the cross section of figure 4 either as a plane in system space (in which case the arrows and flows depict projections, since in general the real flows will point somewhat out of the plane), or as the curved manifold swept out by our one-parameter model as we coarse grain (in which case the flows above our model and below the green curved line should be ignored).

The flow near **S\*** has one unstable direction, leading outward along the green curve (the *unstable manifold*). In system space, there is a surface of points **C** which flow into **S\*** under coarse-graining. Because **S\*** has only one unstable direction, **C** divides system space into two *phases*. To the left of **C**, the systems will have one large, system-spanning avalanche (a snapping noise). To the right of **C**, all avalanches are finite and under coarse-graining they all become small (popping noise). Our model, as it crosses **C** at the value $R_c$, goes through a *phase transition*.

Our model at $R_c$ is not self-similar on the shortest length scales (where the square lattice of domains still is important), but because it flows into **S\*** as we coarse-grain we deduce that it is self-similar on long length scales. Some phase transitions, like ice melting into water, are abrupt and don't exhibit self-similarity. Continuous phase transitions like ours almost always have self-similar fluctuations on long length scales. Also, we must note that our model at $R_c$ will have the *same* self-similar structure as **S\*** does. Indeed, *any* experimental or theoretical model lying on the critical surface **C** will share the same long-wavelength critical behavior. This is the fundamental explanation for universality.

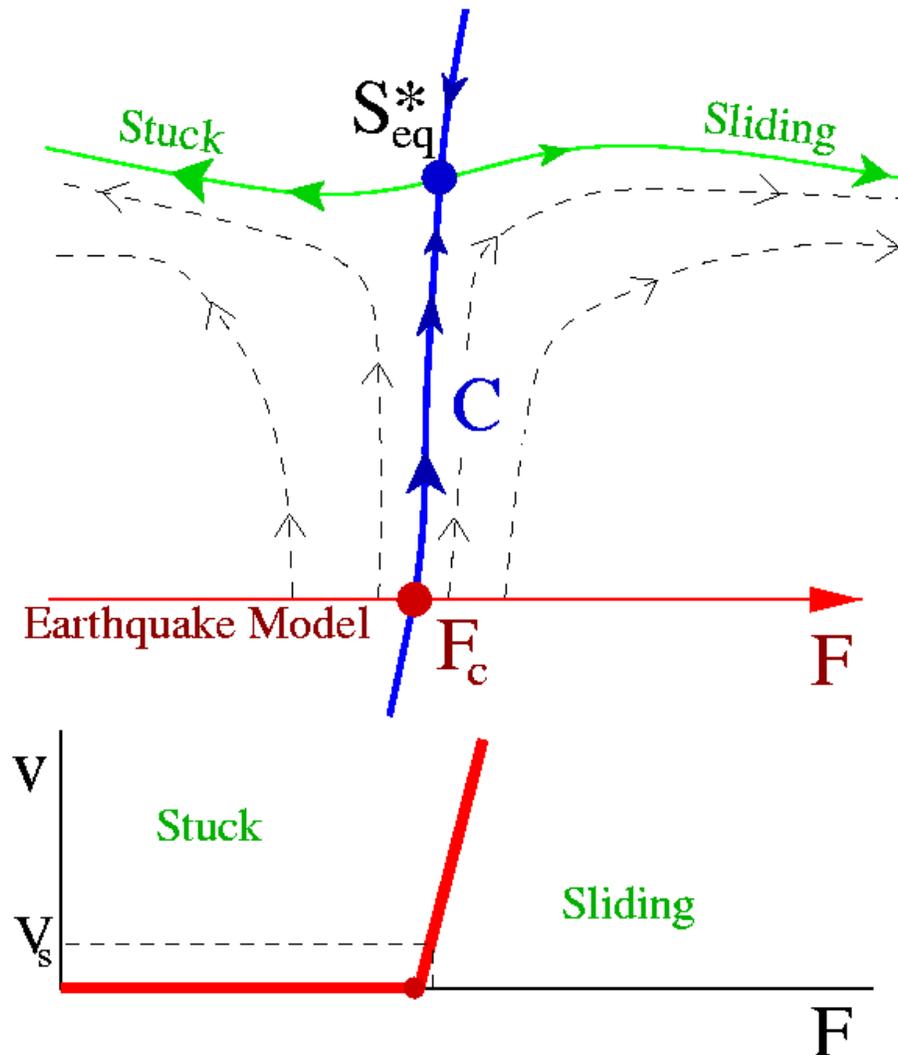

Figure 5: Flows in the space of earthquake models. A model for earthquakes will have a force *F* applied across the front. In models ignoring inertia and velocity-dependent friction[30], there is a critical force $F_c$ that just allows the fault to slip forward. (a) Coarse-graining defines a flow on the space of earthquake models. The fixed point **S\*** will have a different local flow field from other renormalization-group fixed points, yielding its own universality class of critical exponents and scaling functions. The critical manifold **C**, consisting of models which flow into **S\***, separates the stuck faults from those which slide forward with an average velocity *v(F)*. (b) The velocity varies with the external force as a power law $v(F) \sim F^\beta$. The motion of the continental plates, however, does not fix the force *F* across the fault: rather, it sets the average relative velocity to a small value $v_s$ (centimeters per year). This automatically sets the force across the fault very close to its critical force $F_c$. This is one example of *self-organized criticality*.[38,39]

The flows in system space can vary from one class of problems to another: the system space for some earthquake models (figure 5a) will have a different flow, and its fixed point will have different scaling behavior (yielding a different *universality class*). In some cases, a fixed point will attract all the systems in its vicinity (no unstable directions, figure 6). Usually at such attracting fixed points the fluctuations become unimportant at long length scales: the Navier-Stokes equation for fluids described earlier can be viewed as a stable fixed point.[74,75] The coarse-graining process, averaging over many degrees of freedom, naturally smoothens out fluctuations, if they aren't amplified near a critical point by the unstable direction. Fluctuations can remain important when a system has random noise in a conserved property, so that fluctuations can only die away by diffusion: in these cases, the whole phase will have self-similar fluctuations, leading to *generic scale invariance*.[76,77]

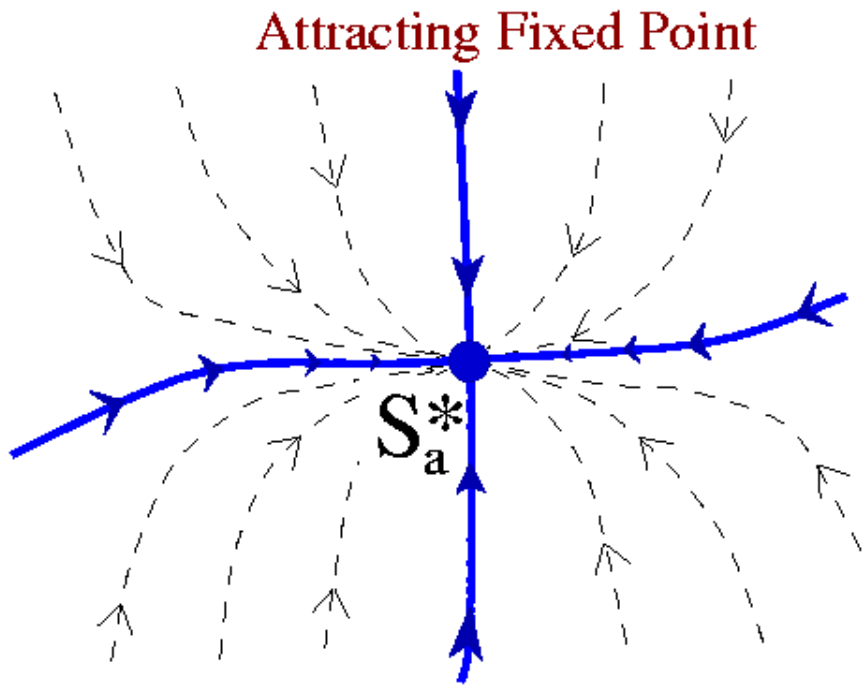

Figure 6: Attracting fixed point. Often there will be fixed points that attract in all directions. These fixed points describe phases rather than phase transitions. Most phases are rather simple, with fluctuations that die away on long length scales.[74] When fluctuations remain important, they will exhibit self-similarity and power laws called *generic scale invariance*.[76,77]

Sometimes, even when the system space has an unstable direction like in figure 4, the observed behavior always has avalanches of all scales. This can occur simply because the physical system averages over a range of model parameters (*i.e.,* averaging over a range of $R$ including $R_c$ in figure 4). For example, this can occur by *the sweeping of a parameter*[78] slowly in time, or varying it gradually in space – either deliberately or through large-scale inhomogeneities.

One can also have *self-organized criticality,*[38,39] where the system is controlled so that it naturally sits on the critical surface. Self-organization to the critical point can occur via many mechanisms. In some models of earthquake faults (figure 5b), the external force naturally stays near the rupture point because the plates move at a fixed, but very small[22] velocity with respect to one another (figure 5b). (This probably does not occur during large earthquakes, where inertial effects lead to temporary strain relief .[30,79]) Sandpile models self-organized (if sand is added to the system at an infinitesimal rate) when open boundary conditions[80] are used (which allows sand to leave until the sandpile slope falls to the critical value). Long-range interactions[81-83] between domains can act as a negative feedback in some models, yielding a net external field that remains at the critical point. For each of these cases, once the critical point is understood adding the mechanism for self-organization is relatively easy.

The case shown in figure 4 of *plain old criticality* is what's seen in some[17,66-69] but not all[81-85] models of magnetic materials, in foams,[44] and in some models of earthquakes.[30]

**Beyond Power Laws**

The renormalization group is the theoretical underpinning for understanding why universality and self-similarity occur. Once we grant that different systems should sometimes share long-distance properties, though, we can quite easily derive some powerful predictions.

To take a tangible example, let's consider the relation between the duration of an avalanche and its size. In paper crumpling, this isn't interesting: all the avalanches seem to be without internal temporal structure.[14] But in magnets large events take longer to finish, and have an interesting internal statistical self-similarity (figure 7a). If we look at all avalanches of a certain duration $T$ in an experiment, they will have a distribution of sizes $S$ around some average $<S>_{experiment}(T)$. If we look at a theoretical model, it will have a corresponding average size $<S>_{theory}(T)$. If our model describes the experiment, these functions must be essentially the same at large $S$ and large $T$. We must allow for the fact that the experimental units of time and size will be different from the ones in our model: the best we can hope for is that $<S>_{experiment}(T) = A <S>_{theory}(T/B)$, for some rescaling factors $A$ and $B$.

Now, instead of comparing to experiment, we can compare our model to itself on a slightly larger time scale.[88] If the time scale is expanded by a small factor $B=1/(1-\delta)$, then the rescaling of the size will also be small, say $1+a\ \delta$.

$<S>(T) = (1 + a\ \delta) <S>( (1-\delta)\ T)$

Making $\delta$ very small yields the simple relation $a <S>=T\ d<S>/dT$, which can be solved to give the power law relation $<S>(T) = S_0\ T^a$. The exponent $a$ is called a *critical exponent*, and is a universal prediction of a given theory. (That means that if the theory correctly describes an experiment, the critical exponents will agree.) In our work, we write the exponent $a$ relating time to size in terms of three other critical exponents, $a=1/\sigma vz$.

### Typical Avalanche Voltage vs. Time
#### Starts and Stops at All Scales

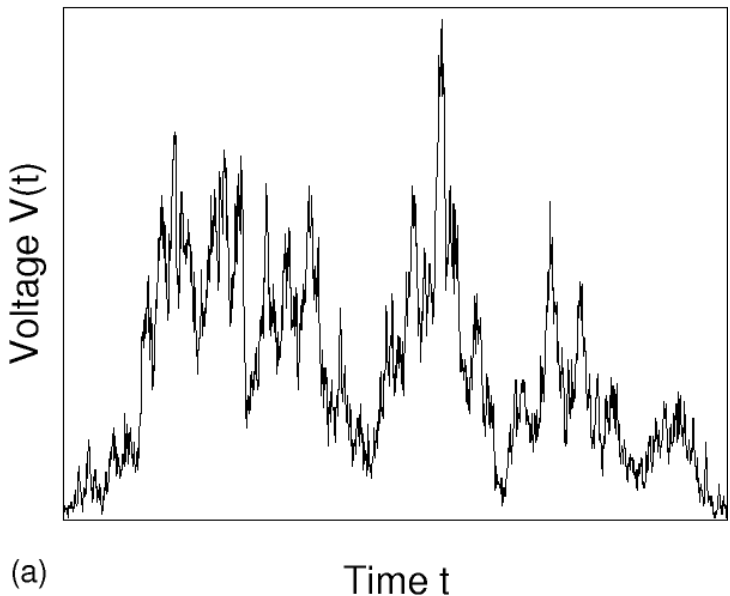

**(a)**

Figure 7: Scaling of avalanche shapes. (a) Voltage (number of domains flipped) pulse during a single large avalanche (arbitrary units). Notice how the avalanche almost stops several times: if the forcing were slightly smaller, this large avalanche would have broken up into two or three smaller ones. The fact that the forcing is just large enough to on average keep the avalanche growing is the cause of the self-similarity: on average a partial avalanche of size $S$ will trigger one other on size $S$. (b) Average avalanche shapes[86] for avalanches of different durations for our model (A. Mehta and K. A. Dahmen). Our theories don't only predict power laws: they should describe all behavior on long length and time scales (at least in a statistical sense). In particular, by fixing parameters one can predict what are called scaling functions. If we average the voltage as a function of time over all avalanches of a fixed duration, we get an average shape. In our simulation, this shape is the same for different durations. (c) Experimental data of Spasojević *et al.*,[87] showing all large avalanches averaged after scaling to fixed duration and area. The experimental average shape is very asymmetric and is not described correctly by our model.

### Average Avalanche Shape: Fixed Duration T
#### Scaling Collapse

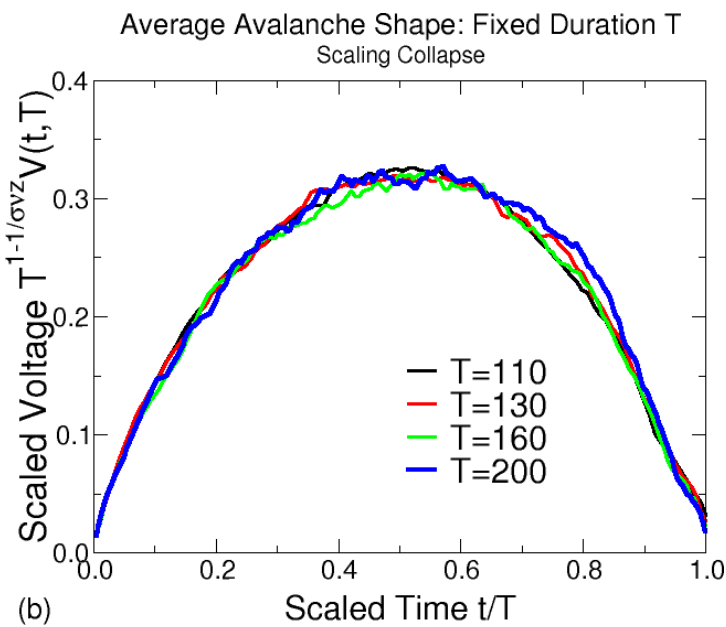

**(b)**

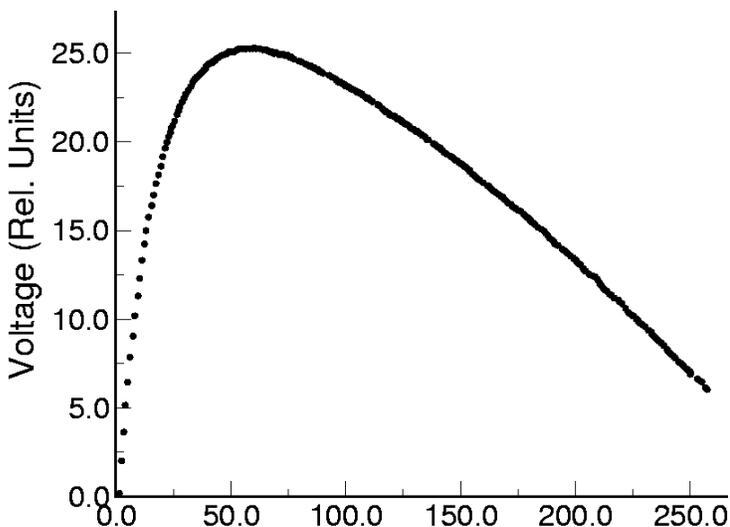

To take a tangible example, let's consider the relation between the duration of an avalanche and its size. In paper crumpling, this isn't interesting: all the avalanches seem to be without internal temporal structure.[14] But in magnets large events take longer to finish, and have an interesting internal statistical self-similarity (figure 7a). If we look at all avalanches of a certain duration $T$ in an experiment, they will have a distribution of sizes $S$ around some average $<S>_{experiment}(T)$. If we look at a theoretical model, it will have a corresponding average size $<S>_{theory}(T)$. If our model describes the experiment, these functions must be essentially the same at large $S$ and large $T$. We must allow for the fact that the experimental units of time and size will be different from the ones in our model: the best we can hope for is that $<S>_{experiment}(T) = A <S>_{theory}(T/B)$, for some rescaling factors $A$ and $B$.

Now, instead of comparing to experiment, we can compare our model to itself on a slightly larger time scale.[88] If the time scale is expanded by a small factor $B=1/(1-\delta)$, then the rescaling of the size will also be small, say $1+a \delta$.

$$<S>(T) = (1 + a \delta) <S>( (1-\delta) T)$$

Making $\delta$ very small yields the simple relation $a <S>=T d<S>/dT$, which can be solved to give the power law relation $<S>(T) = S_0 T^a$. The exponent $a$ is called a *critical exponent*, and is a universal prediction of a given theory. (That means that if the theory correctly describes an experiment, the critical exponents will agree.) In our work, we write the exponent $a$ relating time to size in terms of three other critical exponents, $a=1/\sigma\nu z$.

There are several basic critical exponents, which arise in various combinations depending on the physical property being studied. The details of the naming and relationships between these exponents aren't a focus of our paper. Briefly, the cutoff in the avalanche size distribution in figure 2 gets larger as one approaches the critical disorder as $(R\text{-}R_c)^{-1/\sigma}$ (figure 2). The typical length of the largest avalanche goes as $(R\text{-}R_c)^{-\nu}$. At $R_c$, the probability of having an avalanche of size $S$ goes as $S^{-(\tau+\sigma\beta\delta)}$ (figure 2); if one sits just at the critical field, it goes as $S^{-\tau}$. (Don't confuse the small change in scale $\delta$ with the critical exponent $\delta$.) The fractal dimension of the avalanches is $1/\sigma\nu$, meaning the spatial extent L of an avalanche is proportional to the size $S^{\sigma\nu}$. The duration $T$ of an avalanche of spatial extent L goes as $L^z$…

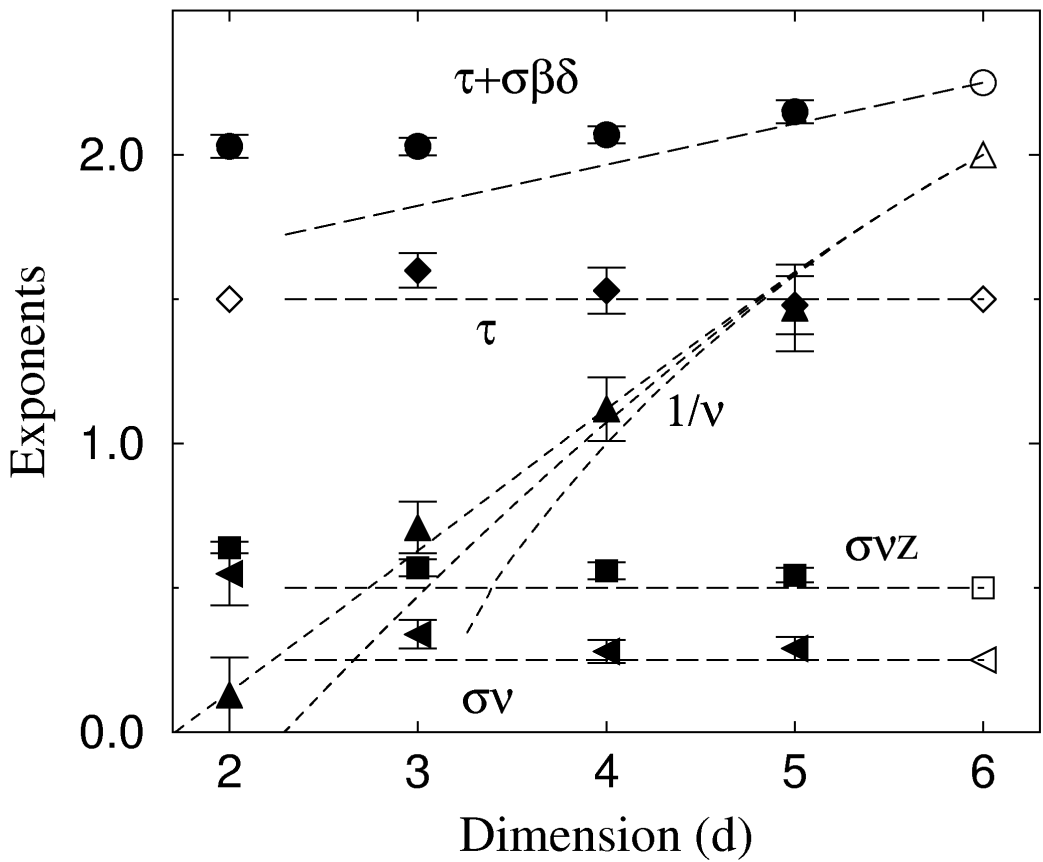

Figure 8: Critical exponents in various dimensions. We test our ε-expansion predictions[72] by measuring[69] the various critical exponents numerically in up to five spatial dimensions. The various exponents are described in the text. All of the exponents are calculated only to linear order in ε, except for the correlation length exponent ν, where we use results from other models.[71,72] The agreement even in three dimensions is remarkably good, considering that we're expanding in ε where ε=3! We should note that perturbing in dimension for our system is not only complicated, but also controversial[89] (see also section VI.C of Ref. 72 and section V of Ref. 69).

To specialists in critical phenomena, these exponents are central; whole conversations will seem to rotate around various combinations of Greek letters. Critical exponents are one of the relatively easy things to calculate from the various analytic approaches, and so have attracted the most attention. They are derived from the eigenvalues of the linearized flows about the fixed point **S\*** in figure 4. Figure 8 shows our numerical estimates[69] for several critical exponents in our model in various spatial dimensions, together with our 6-ε expansions[71,72] for them. Of course the key challenge is not to get analytical work to agree with numerics: it's to get theory to agree with experiment. Figure 9 shows that our model does rather well in describing a wide variety of experiments, but that two rival models (with different flows around their fixed points) also fit.

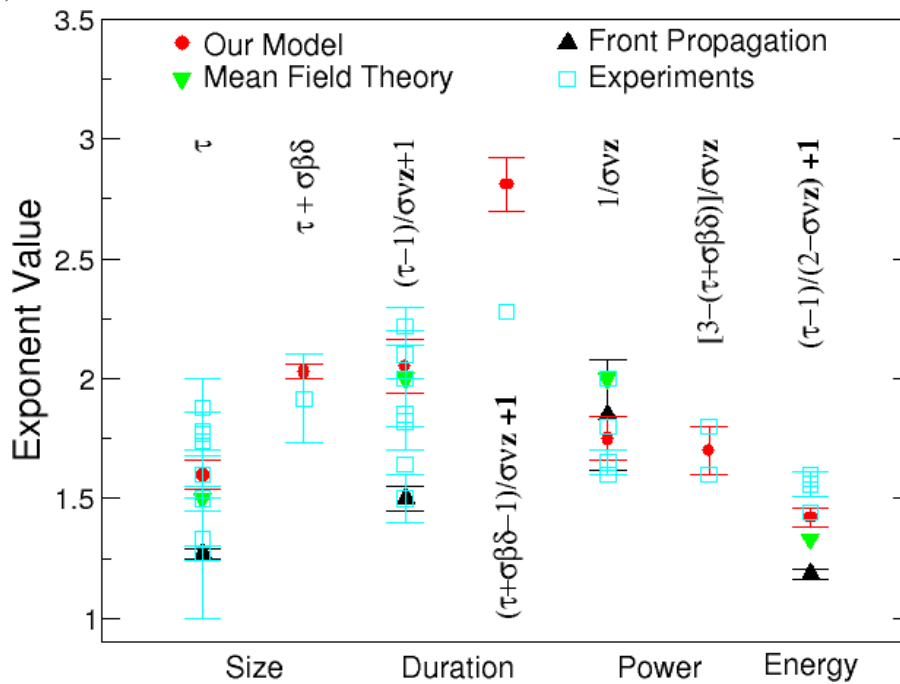

Figure 9: Experiments vs. Theory: Critical Exponents. Different experiments on crackling noise in magnets measure different combinations of the universal critical exponents. Here we compare experimental measurements[81,87,90-101] (see table I of Ref. 66) to the theoretical predictions for three models: our model,[17,66-69] the front-propagation model[48-53,100] and mean-field theory. (In mean-field theory our coupling $J$ in equation [1] couples all pairs of spins: such long-range interactions occur because of the boundaries in models with magnetic dipolar forces.[83] Mean-field theory is equivalent to a rather successful single-degree of freedom model.[102,103]) Power laws giving the probability of getting an avalanche of a given size, duration, or energy at the critical point are shown; also shown is the critical exponent giving the power as a function of frequency[86] (due to the internal structure of the avalanches, Figure 7a). In each pair of columns, the first column includes only avalanches at external fields $H$ in equation [1] where the largest avalanches occur, and the second column (when it exists) includes all avalanches. The various combinations of the basic critical exponents can be derived from exponent equality calculations similar to the one discussed in the text.[66,71,86] Many of the experiments were done years before the theories were developed: many did not report error bars. All three theories do well (especially considering the possible systematic errors in fitting power laws to the experimental measurements: see figure 2). Recent work suggests a clumping of experimental values around the mean-field and front-propagation predictions.[100]

Critical exponents are not the be-all and end-all: many other scaling predictions, explaining wide varieties of behavior, are quite easy to extract from numerical simulations. Universality extends even to those long length scale properties for which one cannot write formulas. Perhaps the most important of these other predictions are the universal scaling functions. For example, let's consider the time history of the avalanches, *V(t)*, denoting the number of domains flipping per unit time. (We call it *V* because it's usually measured as a voltage in a pickup coil.) Each avalanche has large fluctuations, but one can average over many avalanches to get a typical shape. Figure 7b shows the average over all avalanches of fixed duration *T*. Let's call this *<V>(T,t)*. Universality again suggests that this average should be the same for experiment and a successful theory, apart from an overall shift in time and voltage scales:

$<V>_{experiment}(T,t) = A <V>_{theory}(T/B,\ t/B)$. Comparing our model to itself with a shifted time scale becomes simple if we change variables: let *v(T,t/T) = <V>(T,t),* so *v(T,t/T) = A v(T/B,t/T)*. Here *t/T* is a particularly simple example of a *scaling variable*. Now, if we rescale time by a small factor $B=1/(1-\delta)$, we have *v(T, t/T) = (1 + b δ) v(t/T, (1-δ) T)*. Again, making *δ* small we find $b\ v = T\ \partial v/\partial T$, with solution $v = v_0\ T^b$. However, the integration constant $v_0$ will now depend on *t/T*, $v_0 = V(t/T)$, so we arrive at the scaling form

$$<V>\ (t,T) = T^b V(t/T),\qquad\qquad [2]$$

where the entire scaling function *V* is a universal prediction of the theory.

Figures 7b and 7c show the universal scaling functions *V* for our model[86] and an experiment.[87] For our model, we've drawn what are called *scaling collapses,* a simple but powerful way to both check that we're in the scaling regime, and to measure the universal scaling function. Using the form of the scaling equation Eq.[2], we simply plot $T^{-b} <V>(t,T)$ versus *t/T*, for a series of long times *T*. All the plots fall onto the same curve. This tells us that our avalanches are large enough to be self-similar. (If in your scaling collapse the corresponding plots do not all look alike, then any power laws you have measured are probably accidental.) It also provides us with a numerical evaluation of the scaling function $V$. Note that we use $1/\sigma vz\text{-}1$ for the critical exponent $b$. This is an example of an *exponent equality*: easily derived from the fact that $<S>(T) = \int <V>(t,T)\ dt = \int T^b\ V(t/T)\ dt \sim T^{b+1}$, and the scaling relation $<S>(T) \sim T^{1/\sigma vz}$.

Notice that our model and the experiment have quite different shapes for $V$. The other two models from Figure 9 also give much more symmetrical forms for $V$ than the experiment does.[86] How do we react to this? Our models are falsified if any of the predictions are shown to be wrong asymptotically on long length and time scales. If duplication of this measurement by other groups continues to show this asymmetry, then our theory is obviously incomplete. Even if later experiments in other systems agree with our predictions, it would seem that this particular system is described by an undiscovered universality class. Incorporating insights from careful experiments to refine the theoretical models has historically been crucial in the broad field of critical phenomena. The message we emphasize here is that scaling functions can provide a sharper tool for discriminating between different universality classes than critical exponents.

Broadly speaking, most common properties that involve large length and time scales have scaling forms: using self-similarity, one can write functions of $N$ variables in terms of scaling functions of $N\text{-}1$ variables: $F(x,y,z) = z^{-\alpha}\ F(x/z^\beta,\ y/z^\gamma)$. In the inset to figure 2, we show the scaling collapse for the avalanche size distribution: $D(S,R) = S^{-(\tau+\sigma\beta\delta)}\ D(\ (R\text{-}R_c)\ /\ S^{-\sigma})$. (This example illustrates that scaling works not only at $R_c$ but also near $R_c$; the green unstable manifold in figure 4 governs the behavior for systems near the critical manifold $\mathbf{C}$.)

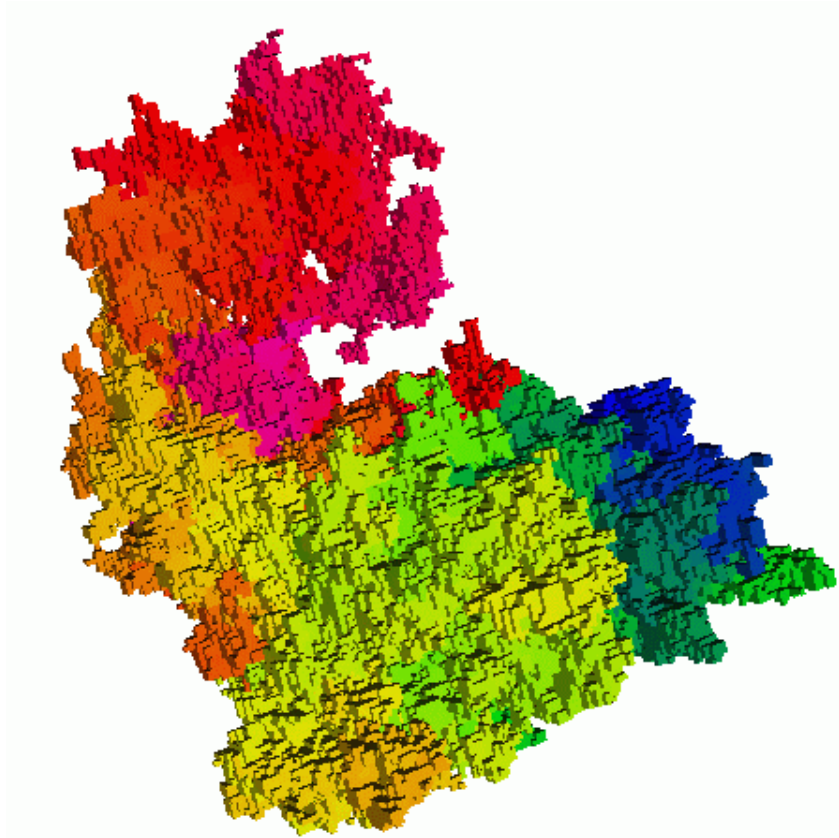

Figure 10: Fractal spatial structure of an avalanche.[67] Fractal structures, as well as power laws, are characteristic of systems at their critical point. This moderate-sized avalanche involved the flipping of 282,785 domains in our simulation. The colors represent time: the first domains to flip are colored blue, and the last pink. So far, there have not been many experiments showing the spatial structure of avalanches.[104] When experiments become available, there are a wealth of predictions of the scaling theories that we could test. Other systems[105,55,30] have seen a qualitatively different kind of avalanche spatial structure, where the avalanche is made up of many small disconnected pieces, which trigger one another through the waves emitted as they flip.

Many other kinds of properties beyond critical exponents and scaling functions can be predicted from these theories. Figure 10 shows the spatial structure of a large avalanche in our model: notice not only that it is fractal (rugged on all scales), but that it is longer than it is wide,[106] and that it is topologically interesting.[107] (It has tunnels, and sometimes during the avalanche it forms a tunnel and later winds itself through it, forming a knot. It's interesting that the topology of the interfaces in the three-dimensional Ising model have applications in quantum gravity.[107]) The statistics of all of these properties, in other systems, have been shown to be universal on long length and time scales.

**Continuing challenges**

Do not be fooled: our understanding of crackling noise is far from complete. There are only a few systems[30,34,50,51,54.2,54.4,71-73] where the renormalization-group framework has substantially explained even the behavior of the numerical models. There are several other approaches[63,64,80,108-110] that have been developed to study crackling noise, many which share our view of developing effective descriptions on long length and time scales. But the successes remain dwarfed by the bewildering variety of systems that crackle. Getting a global perspective on the universality classes for crackling noise remains an open challenge.

An even more important challenge is to make quantitative comparison between the theoretical models and experimental systems. We believe that going beyond power laws will be crucial in this endeavor. The past focus on critical exponents has sometimes been frustrating: it's too easy to find power laws over limited scaling ranges,[111] and too easy to find models which roughly fit them. It also seems unfulfilling, summarizing a complex morphology into a single critical exponent. We believe that measuring a power law is almost never definitive by itself: a power law in conjunction with evidence that the morphology is scale invariant (*e.g.,* a scaling collapse) is crucial. By aggressively pursuing quantitative comparisons of other, richer measures of morphology such as the universal scaling functions, we will be better able both to discriminate among theories and to ensure that a measured power-law corresponds to a genuine scaling behavior.

Another challenge is to start thinking about the key ways that these complex spatiotemporal systems *differ* from the phase transitions we understand from equilibrium critical behavior. (The renormalization-group tools developed by our predecessors are seductively illuminating: it's always easy to focus where the light is good.) For example, in several of these systems there are collective, dynamical

"memory" effects[17,112-115] that even may have practical applications.[116] The quest for a scaling theory of crackling phenomena needs to be viewed as part of the larger process of understanding the dynamics of these nonlinear, nonequilibrium systems.

A final challenge is to make the study of crackling noise profitable. Less noise from candy wrappers[14-16] in movie theaters is not the most pressing of global concerns. Making money from fluctuations in stock prices is already big business.[58,59] Predicting earthquakes over the short term probably will not be feasible using these approaches,[117] but longer term prediction of impending large earthquakes may be both possible[79] and useful, say, for guiding local building codes. Understanding that the large-scale behavior relies only on a few emergent material parameters (disorder and external field for our model of magnetism) leads one to study how these parameters depend on the microphysics: one has dreams, for example, of learning how to shift an active earthquake fault into a harmless, continuously sliding regime by adding lubricants to the fault gouge. In the meantime, crackling noise is basic research at its elegant, fundamental best.

## Acknowledgements

Much of the work reviewed in this paper grew out of a collaboration with Matt Kuntz[2]. We thank Amit Mehta for supplying the data for Figure 7b. We'd like to thank Drew Dolgert, Mark Newman, Jean-Phillipe Bouchaud, L. Cindy Krysac, Daniel Fisher, and Jim Thorpe for helpful comments and references.

This work was supported by NSF grants DMR 9805422, 9873214, 00-72783, 99-76550, ASC-9523481, the Cornell Theory Center (through NSF #9972853) and IBM.